\documentclass[twocolumn]{aastex63}
\newcommand{\HI}{\ion{H}{1}}
\newcommand{\NII}{\ion{N}{2}}
\newcommand{\HII}{\ion{H}{2}}
\newcommand{\kms}{\mbox{km~s$^{-1}$}}
\newcommand{\Msol}{\mbox{M$_\odot$}}
\newcommand{\surm}{\mbox{M$_\odot$ pc$^{-2}$}}

\newcommand{\siggas}{\mbox{$\Sigma_{\rm gas}$}}
\newcommand{\sigsfr}{\mbox{$\Sigma_{\rm SFR}$}}

\newcommand{\sighi}{\mbox{$\Sigma_{\rm HI}$}}
\newcommand{\sightwo}{\mbox{$\Sigma_{\rm H_2}$}}

\newcommand{\coj}{\mbox{$^{12}$CO ($J=1\rightarrow0$)}}
\newcommand{\um}{\mbox{$\micron$}}
\newcommand{\ac}{\mbox{$\arcsec$}}
\newcommand{\Htwo}{\mbox{H$_2$}}
\newcommand{\rhohtwo}{\mbox{$\rho_{\rm H_2}$}}
\newcommand{\rhohi}{\mbox{$\rho_{\rm HI}$}}
\newcommand{\rhosfr}{\mbox{$\rho_{\rm SFR}$}}
\newcommand{\rhogas}{\mbox{$\rho_{\rm gas}$}}

\received{2022 June 27 }
\accepted{2022 October 7}

\shorttitle{The volumetric star formation law in NGC 4302}
\shortauthors{Yim et al.}

\begin{document}

\title{The Volumetric Star Formation Law in the Almost Edge-on Galaxy NGC 4302 Revealed by ALMA}

\correspondingauthor{Kijeong Yim}
\email{kijeong.yim@gmail.com}

\author[0000-0002-3426-5854]{Kijeong Yim}
\affiliation{Department of Astronomy and Space Science, Chungnam National University, Daejeon 34134, Republic of Korea}
\affiliation{Korea Astronomy and Space Science Institute, 776 Daedeok-daero, Yuseong-gu, Daejeon 34055, Republic of Korea}

\author[0000-0002-7759-0585]{Tony Wong}
\affiliation{Department of Astronomy, University of Illinois, 1002 West Green Street, Urbana, IL 61801, USA}

\author[0000-0003-2048-4228]{Richard J. Rand}
\affiliation{Department of Physics and Astronomy, University of New Mexico, 1919 Lomas Blvd NE, Albuquerque, NM 87131-1156, USA}





\begin{abstract}
We observe the almost edge-on ($i \sim 90\degr$) galaxy NGC 4302 using ALMA (CO) and VLA (\HI) to measure the gas disk thickness for investigating the volumetric star formation law (SFL). The recent star formation rate (SFR) is estimated based on a linear combination of IR 24 \um\ and H$\alpha$ emissions.
The measured scale heights of CO and \HI\ increase significantly with radius. Using the scale heights along with the vertically integrated surface densities, we derive the mid-plane volume densities of the gas (\rhohi, \rhohtwo, and \rhogas = \rhohi+\rhohtwo) and the SFR (\rhosfr) and compare the volumetric SFL (\rhosfr $\propto$ \rhogas$^{n}$) with the vertically integrated SFL (\sigsfr $\propto$ \siggas$^{N}$). We find tight power-law correlations between the SFR and the gas (\HI, \Htwo, and the total gas) in both volume and surface densities. The power-law indices of the total gas and \HI\ for the volumetric SFL are noticeably smaller than the indices for the vertically integrated SFL while the \Htwo\ indices for both cases are similar to each other. In terms of the star formation efficiency (SFE), we find that the molecular and total gas SFEs are roughly constant, while the atomic SFE is clearly decreasing with radius in both cases.
\end{abstract}

\keywords{Interstellar medium (847); Barred spiral galaxies (136); Star formation (1569); Galaxy structure (622); Galaxy kinematics (602); Galaxy dynamics (591); Radio astronomy (1338)}




\section{Introduction} \label{sec:intro}
A power-law correlation between the SFR (\sigsfr) and gas (e.g., \sightwo, \sighi, \siggas) based on surface densities in nearby galaxies is reported by a number of observational studies (e.g., \citealt{2002ApJ...569..157W}; \citealt{2008AJ....136.2846B}; \citealt{2011AJ....142...37S}; \citealt{2014AJ....148..127Y}, \citeyear{2016MNRAS.463.2092Y}) following the work by \citet{1998ApJ...498..541K}, who demonstrated the correlation between \sigsfr\ and \siggas\ (=\sightwo+\sighi) averaged over the disks: \sigsfr\ $\propto \Sigma_{\rm gas}^{1.4}$. However, the volume (not surface) density is the relevant quantity for the Jeans instability and the free-fall time scale, and indeed the original \citet{1959ApJ...129..243S} star formation law was based on volume density. \citet{2012ApJ...745...69K} showed that the volumetric star formation law is well matched to observational data including Milky Way molecular clouds, Local Group galaxies, and high-redshift galaxies. Recently, \citet{Bacchini_2019} investigated the relationship between gas and the SFR volume densities estimated from  hydrostatic equilibrium and found a strong correlation between them with a smaller scatter compared to the vertically integrated star formation law (Kennicutt-Schmidt law) in nearby disk galaxies. \citet{2020A&A...644A.125B} suggested that the volumetric SFL is valid even in dwarf galaxies and outer regions of spiral galaxies where the vertically integrated SFL appears to break down due to \HI\ predominance and low metallicity in the regions. In our previous study \citep{2020MNRAS.494.4558Y}, we investigated the volumetric SFL by measuring the gas disk thickness directly from the edge-on galaxies observed by the Combined Array for Research in Millimeter-wave Astronomy (CARMA) in contrast to volumetric SFL studies using the inferred disk thickness from hydrostatic equilibrium (e.g., \citealt{2008ARep...52..257A}; \citealt{Bacchini_2019}, \citeyear{2020A&A...644A.125B}). However, the CO disk thickness is not well-resolved by CARMA and the angular resolution is about 3\ac, corresponding to a physical scale of $\sim$200 pc in the edge-on galaxy sample. 
For this reason, we use the Atacama Large Millimeter/submillimeter Array (ALMA), providing unprecedented sensitivity and resolution, to resolve the CO disk structure for studying the volumetric SFL. 

NGC 4302 is an excellent target, allowing us to measure the disk thickness in the most straightforward way since its inclination is almost $90\degr$ (\citealt{Heald_2007}; \citealt{Zschaechner_2015}). It is located at a distance of 14 Mpc based on a local flow model by \cite{2000ApJ...529..786M} with $H_0$ = 73 \kms\ Mpc$^{-1}$. NGC 4302 has a companion galaxy NGC 4298, but the disk of NGC 4302 is not obviously distorted by the interaction (\citealt{1996ApJ...462..712R}; \citealt{Howk_1999}). 
On the other hand, \citet{Zschaechner_2015} detected a faint \HI\ bridge between NGC 4302 and NGC 4298 in a few channel maps, probably due to ram pressure stripping by the intracluster medium in the Virgo Cluster.


\begin{figure*}
\begin{center}
\includegraphics[width=0.9\textwidth]{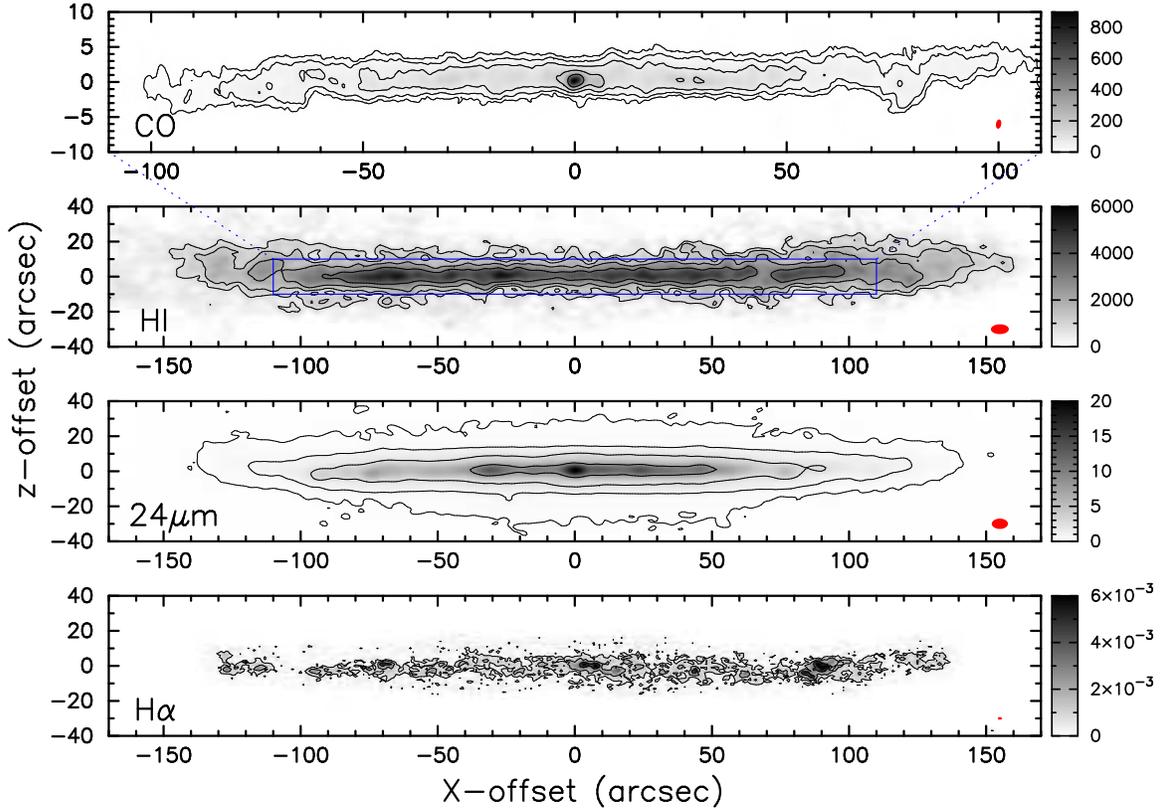}\\
\caption{CO and \HI\ integrated intensity maps. The northern disk is placed in negative $x$. Contour levels are $10.0 \times 2.5^n$ K \kms, with n=0, 1, 2, 3, 4 for CO and $750.0 \times 1.6^n$ K \kms, with n=0, 1, 2, 3 for \HI. The lowest contour levels are $\sim$10$\sigma$ for CO and $\sim$5$\sigma$ for \HI. The synthesized beam is shown in the lower right corner of each panel. The $Spitzer$ 24 \um\ contour levels are 0.2 $\times$ 3.4$^n$ MJy sr$^{-1}$, with n = 0, 1, 2, and 3 and the H$\alpha$ contour levels are 0.0003 $\times$ 2.5$^n$ MJy sr$^{-1}$, with n = 0, 1, and 2. The lowest contours are $\sim$10$\sigma$ for 24 \um\ and $\sim$3$\sigma$ for H$\alpha$.
The point spread function (PSF) FWHM is shown in the lower right corner of each panel: 5.9\ac\ for 24 \um\ and 1.5\ac\ for H$\alpha$.  
\label{fig:map}}
\end{center}
\end{figure*}

\section{Observations and Data Reduction} \label{sec:obs}
We observed \coj\ emission toward NGC 4302 using ALMA (Cycle 7) 12m in array configurations C43-4 (6.7 hours) and C43-1 (1.9 hours) with a twenty-five point mosaic in 2019 October 20th -- 2020 March 4th. Using a pipeline script of the Common Astronomy Software Applications (CASA) package \citep{2007ASPC..376..127M} provided by ALMA, we calibrated separately the visibility data in different configurations. The calibrated data were combined using the CASA task CONCAT and imaged using the CASA task TCLEAN with Briggs weighting (ROBUST = 0.5). A primary beam correction was applied to the cleaned image using IMPBCOR. 
The combined image has an angular resolution of 1$\farcs$42 $\times$ 1$\farcs$16 (P.A. = $-39.2\degr$) and a velocity resolution of 5 \kms. The rms noise per channel is $\sim$1.3 mJy Beam$^{-1}$ ($\sim$72.6 mK). Figure \ref{fig:map} (top) shows the integrated intensity map of the CO cube. In order to place the redshifted side (southern disk) on the right,  the map is rotated by 90.93$\degr$ based on a position angle of -0.93$\degr$ that we measured from $Spitzer$ 3.6 \um\ image using the MIRIAD task IMFIT. The galactic center ($\alpha$ = \rm{12$^h$21$^m$42$\fs$32 and $\delta$ = 14$\degr$35$\arcmin$52.23\ac}) is located at $x = 0$ and $z = 0$ in the map. We masked the CO cube to reduce noise in the integrated intensity map by blanking regions below a 3$\sigma$ threshold in a cube smoothed to FWHM = 3\ac. The total flux of the integrated intensity map is $\sim$650 Jy \kms. The sensitivity falls to 50\% at $x$-offset of $\sim$120\ac. 
In order to examine the flux recovery, we compare the ALMA observations with 45 m single-dish NRO (Nobeyama Radio Observatory) observations \citep{Komugi_2008} in the same region over the velocity range of 990--1170 \kms, chosen by \cite{Komugi_2008} to calculate the integrated intensity. The ALMA data are convolved to the circular beam of 16\ac, which is the angular resolution of the NRO observations. The integrated intensities over the same region from 990 \kms\ to 1170 \kms are 25 K \kms\ for NRO and 16 K \kms\ for ALMA. This indicates that the ALMA observations recover about 64\% of the flux.


The \HI\ observations were carried out with VLA in B configuration in 2009 \citep{Zschaechner_2015} and C configuration in 2005 \citep{Chung_2009}. 
We used the CASA package tasks FLAGCMD (mode=`manual') for flagging bad data, SETJY for setting the flux density of the amplitude calibrator, BANDPASS for the bandpass correction, GAINCAL for the gain calibration, FLUXSCALE for bootstrapping the flux density of the gain calibrator, APPLYCAL for applying the calibration solutions to the target, and CONCAT for combining the separately calibrated B and C data. 
Finally, we cleaned the combined data using the task CLEAN and subtracted continuum emission from the image using the task IMCONTSUB. In the cleaning process, we used Briggs weighting with a robustness of 0.5, which resulted in an angular resolution of 6$\farcs$56 $\times$ 5$\farcs$59 (P.A. = $-87.4\degr$).  
The masked integrated intensity map of the \HI\ cube, consisting from 940 \kms\ to 1340 \kms\ with a channel width of 20 \kms, is shown in  Figure \ref{fig:map} (middle). The mask was created from the 3$\sigma$ threshold of a cube smoothed to 15\ac. The rms noise per channel is $\sim$2.3 K and the total flux of the integrated intensity map is 28 Jy \kms. For comparison, the integrated flux density from the Arecibo Legacy Fast ALFA (ALFALFA) survey is 26.81 Jy \kms\  \citep{2007AJ....133.2569G}. The single dish flux is in good agreement with our measurement within uncertainties.



In order to estimate the SFR, we used both 24 \um\ and H$\alpha$ emissions from dust and  \HII\ regions surrounding young massive stars, respectively.
For the $Spitzer$ MIPS 24 \um\ (Program ID 30945; PI: Jeffrey Kenney), we downloaded the basic calibrated data (BCD) from the Spitzer Heritage Archive and removed the residual artifacts such as latents and gradients (MIPS Instrument Handbook\footnote {https://irsa.ipac.caltech.edu/data/SPITZER/docs/mips/}) in the BCD via Image Reduction and Analysis Facility (IRAF) tasks. After mosaicking the BCD images using MOPEX (Mosaicking and Point Source Extraction), we employed the Groningen Image Processing System (GIPSY; \citealt{1992ASPC...25..131V}) task BLOT to mask out bright stars of the 24 \um\ image in Figure \ref{fig:map} (third panel). The H$\alpha$ image in the bottom panel was observed by \cite{1996ApJ...462..712R} with the 2.1 m Kitt Peak National Observatory (KPNO). The continuum subtraction was carried out in the image and there is no contamination by the [\NII] lines \citep{1996ApJ...462..712R}. 


\begin{figure}
\begin{center}
\includegraphics[width=0.45\textwidth]{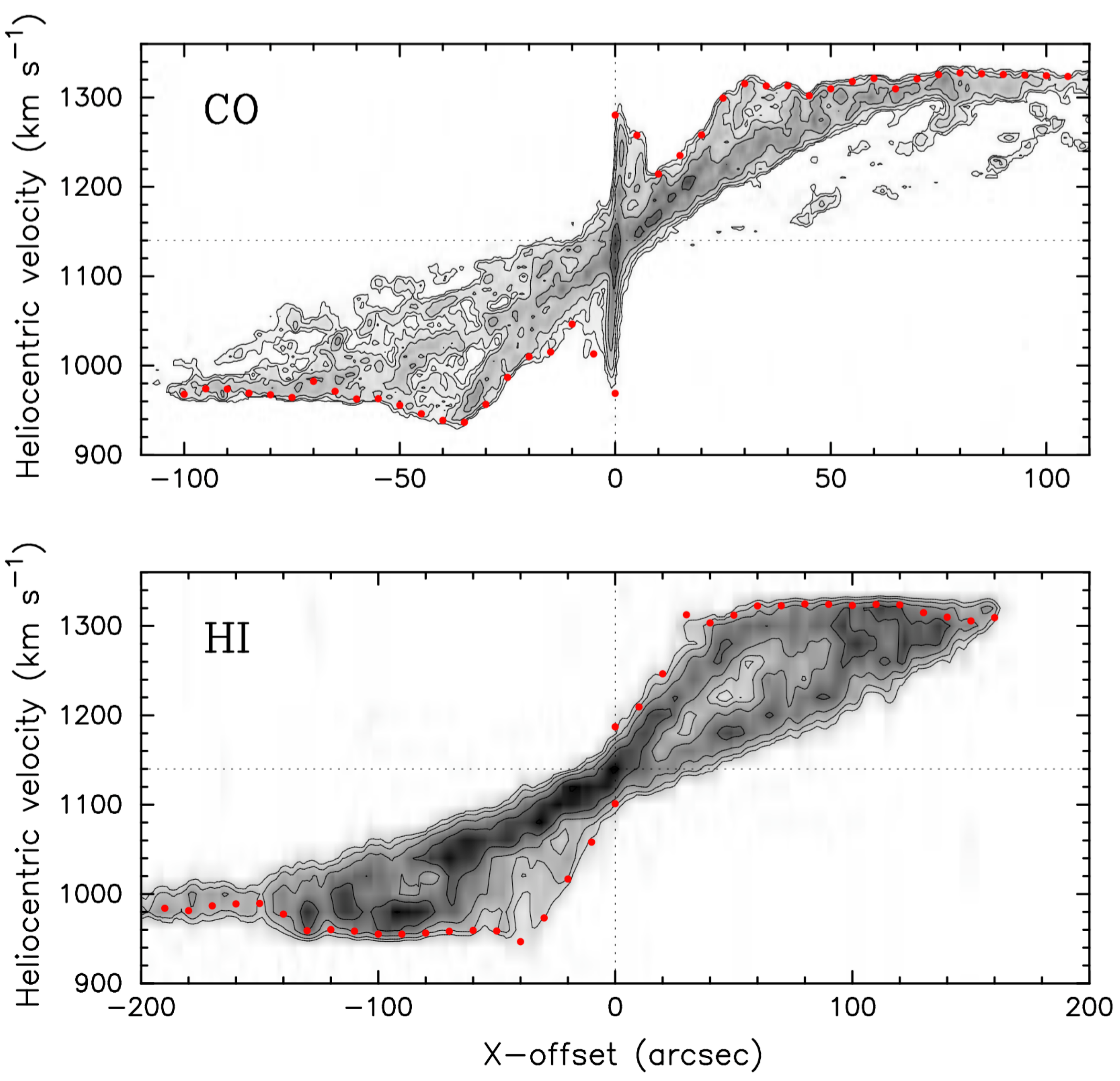}\\
\includegraphics[width=0.4\textwidth]{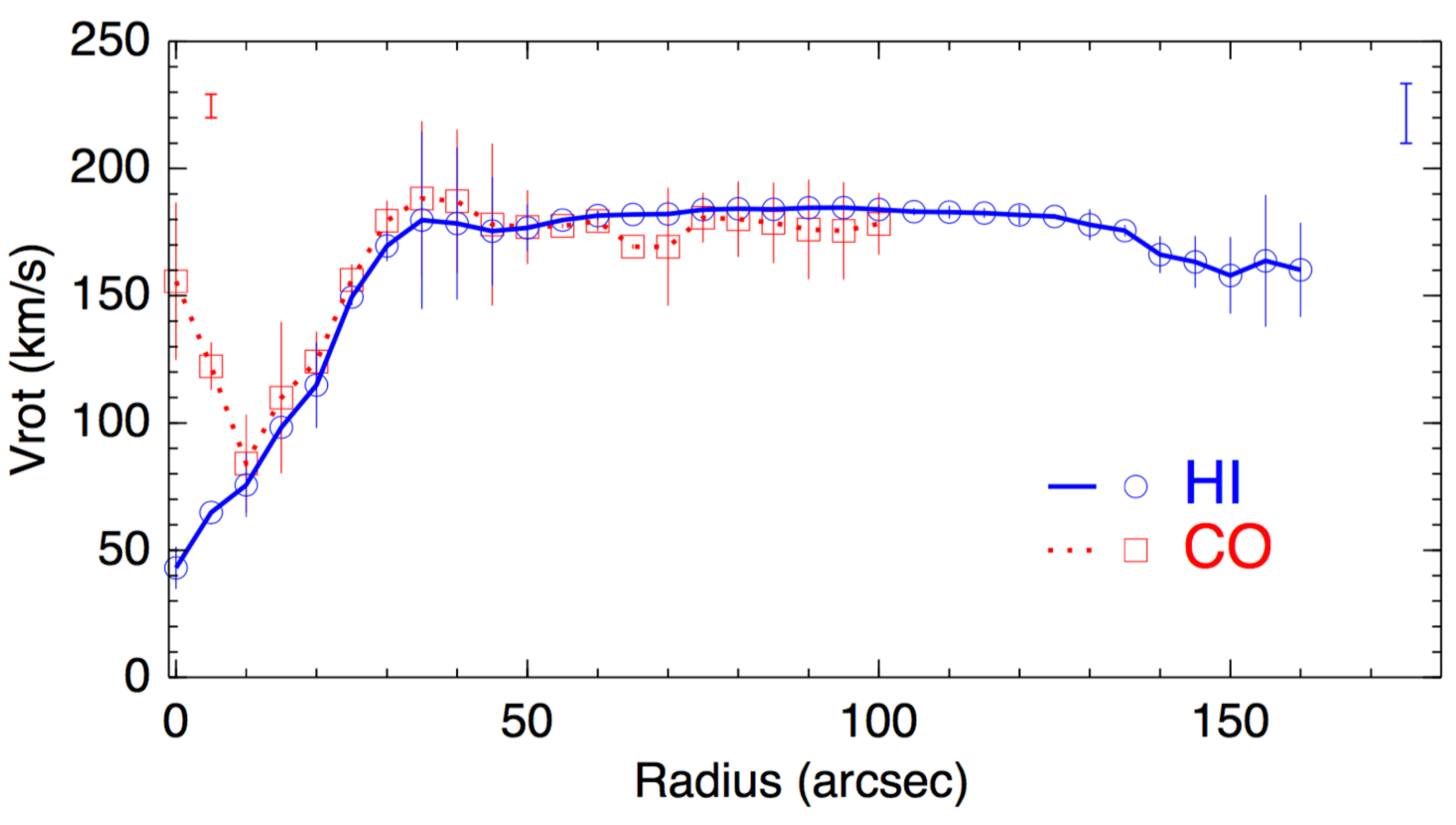}
\caption{Position-velocity diagrams integrated vertically for CO (top) and \HI\ (middle). CO contours are $1.0 \times 1.8^n$ K arcsec, with n=0, 1, 2, 3, 4.
\HI\ contour levels are $175.0 \times 1.5^n$ K arcsec, with n=0, 1, 2, 3, 4.  The lowest contour level is 5$\sigma$. The horizontal dotted lines indicate the heliocentric systemic velocity of 1140 \kms. Bottom: Rotation curves of CO (red squares) and \HI\ (blue circles) obtained from p-v diagrams along the mid-plane. The rotation velocities are overlaid on the p-v diagrams as red filled circles. The uncertainties from the correction term in Equation \ref{eq_vrot} are shown in the upper left and right corners.
\label{fig:pvd}}
\end{center}
\end{figure}

\section{Kinematics} \label{sec:kin}

\subsection{Position-Velocity Diagrams}
The position-velocity (p-v) diagram allows us to obtain the rotation curve using the envelope tracing method which uses the terminal velocity corrected for the instrumental resolution and gas velocity dispersion (\citealt{2001ARA&A..39..137S}; see Section 3.3) and the radial density distribution based on the PVD method which assumes axisymmetry (\citealt{2011AJ....141...48Y}; see Section 4.1). We obtained the p-v diagrams (Figure \ref{fig:pvd}) by integrating the data cubes over $\pm$10\ac\ for CO and $\pm$30\ac\ for \HI\ from the mid-plane. The horizontal dotted line indicates the systemic velocity   $V_{\rm sys}$ of 1140 \kms\ that we adopted based on both the consistency of the redshifted and blueshifted rotation curves and HyperLEDA\footnote {http://leda.univ-lyon1.fr} \citep{2014A&A...570A..13M}.

The p-v diagram of CO (top panel of Figure \ref{fig:pvd}) shows a central elongated feature followed by a gap, possibly caused by a bar such as in NGC 891  \citep{2011AJ....141...48Y} and NGC 4013 \citep{2020MNRAS.494.4558Y}. The maximum radial velocity of the central feature ($\sim$1280 \kms) is lower than the almost flat velocity ($\sim$1320 \kms) beyond $\sim$30\ac\ in the redshifted side. The lower velocity compared to the flat velocity suggests an end-on bar (e.g. NGC 4013) while a higher velocity in the central feature suggests a side-on bar as in NGC 891 when we are seeing the $x_2$ periodic orbits in the center \citep{1999ApJ...522..699A}. Therefore, we suggest that a bar of NGC 4302 is located close to the line of sight (i.e. end-on bar). Unlike the CO p-v diagram, the elongated feature in the central region does not appear in the \HI\ p-v diagram (middle panel of Figure \ref{fig:pvd}); this is perhaps because the central region has little \HI\ emission. The p-v diagram clearly shows that the emission in the blueshifted side is much further extended than the emission in the redshifted side. 
\citet{Chung_2007} reported the extended emission as a long one-sided \HI\ tail and they suggest that the \HI\ tail could be caused by ram pressure stripping.

\subsection{Rotation Curve}
We derived the CO and \HI\ rotation curves using the p-v diagrams along the mid-plane through the envelope tracing method. This method calculates the rotation velocities ($V_{\rm rot}$) based on the terminal velocities ($V_{\rm ter}$) of the p-v diagram, the observational velocity resolution ($\sigma_{\rm obs}$) and the velocity dispersion of gas ($\sigma_{\rm g}$):
\begin{equation}
V_{\rm rot} = V_{\rm ter} - \sqrt{\sigma^2_{\rm obs} + \sigma^2_{\rm g}},
\label{eq_vrot}
\end{equation}
where $\sigma^2_{\rm obs}$ for CO and \HI\ is 2.2 \kms\ and 8.5 \kms\ based on the channel resolutions, respectively. 
The adopted velocity dispersions of CO and \HI\ are 4 \kms\ (\citealt{2011MNRAS.410.1409W}; \citealt{2016AJ....151...15M}; \citealt{2017A&A...607A.106M}) and 8 \kms\ (\citealt{1984A&A...132...20S}; \citealt{2006ApJ...650..933B}), respectively. The adopted values do not noticeably affect the results.   
The terminal velocity is defined as the highest velocity at the 3$\sigma$ contour level on the mid-plane p-v diagram. The obtained rotation velocities are shown as red circles overlaid on the integrated p-v diagrams in Figure \ref{fig:pvd}. The rotation curve is very uncertain near the center due to the existence of a bar. The blueshifted side is in good agreement with the redshifted side for both CO and \HI. The rotation curves in the bottom panel of Figure \ref{fig:pvd} show average values of the blueshifted and redshifted velocities as red open circles for CO and blue crosses for \HI. The differences between the blueshifted and redshfited velocities are used as the error bars on the averages. In addition, the uncertainties from the correction term ($\sqrt{\sigma^2_{\rm obs} + \sigma^2_{\rm g}}$) in Equation \ref{eq_vrot} are shown in the upper left and right corners in the figure.

\section{Radial Distribution}
The density distributions as a function of radius are necessary to  compare the gas with the SFR for investigating the star formation law. We obtained the radial profiles using the PVD method for the gas and  the GIPSY task RADPROF for the SFR. The CO, \HI, 24 \um, and H$\alpha$ maps are convolved to a circular beam of 6.56\ac\ $\times$ 6.56\ac\ based on the beam size of \HI\ before deriving the radial profiles in order to compare and/or combine them. When we convolve the 24 \um\ and H$\alpha$ maps, we use the kernels provided by \citet{2011PASP..123.1218A}.
The assumed original point spread functions for 24 \um\ and H$\alpha$ are 5.9\ac\ and 1.5\ac\ \citep{1996ApJ...462..712R}, respectively. 

\subsection{Molecular and Atomic Gas}
\label{sec:radial}

\begin{figure}
\begin{center}
\includegraphics[width=0.35\textwidth]{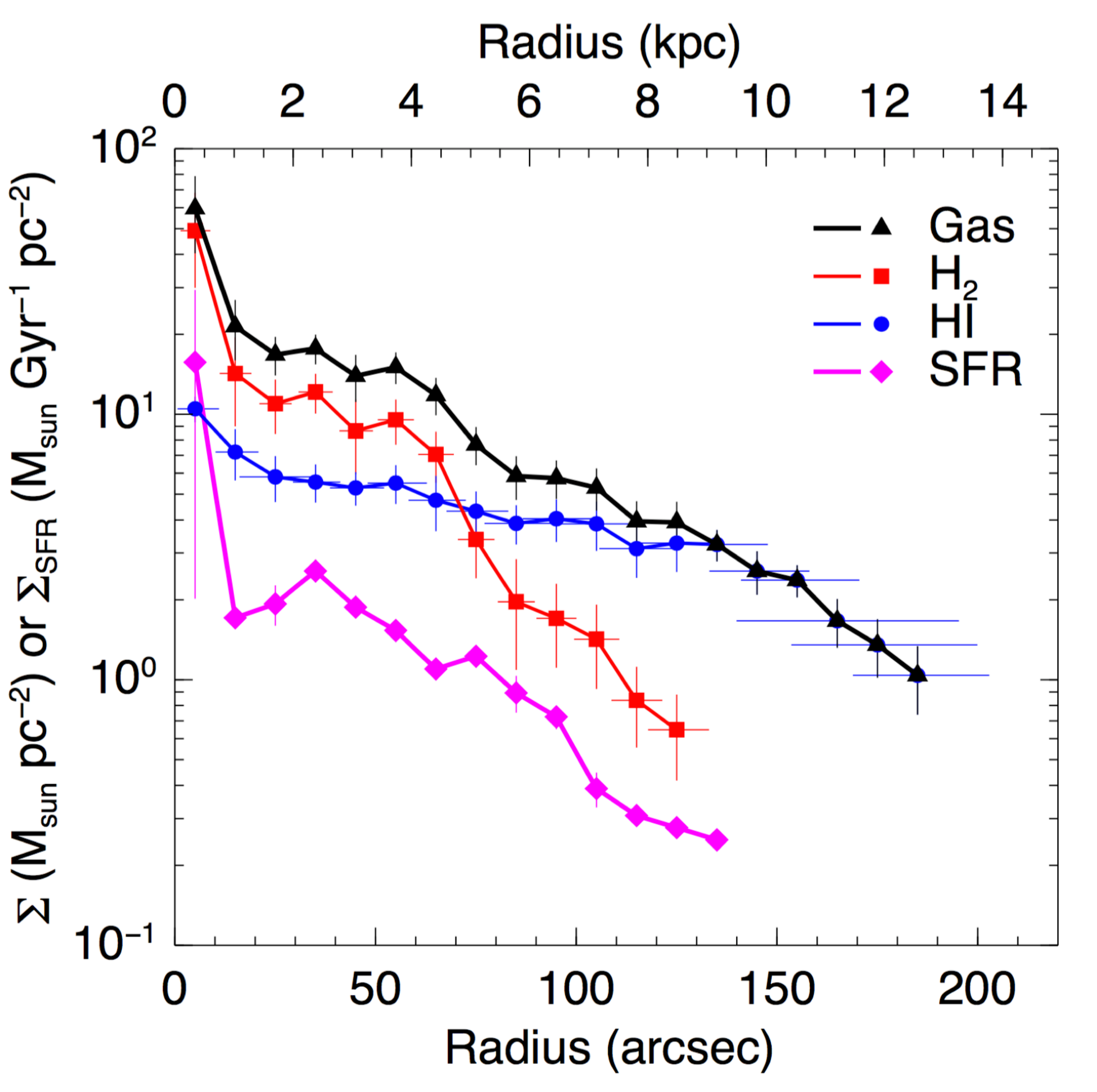}\\
\includegraphics[width=0.35\textwidth]{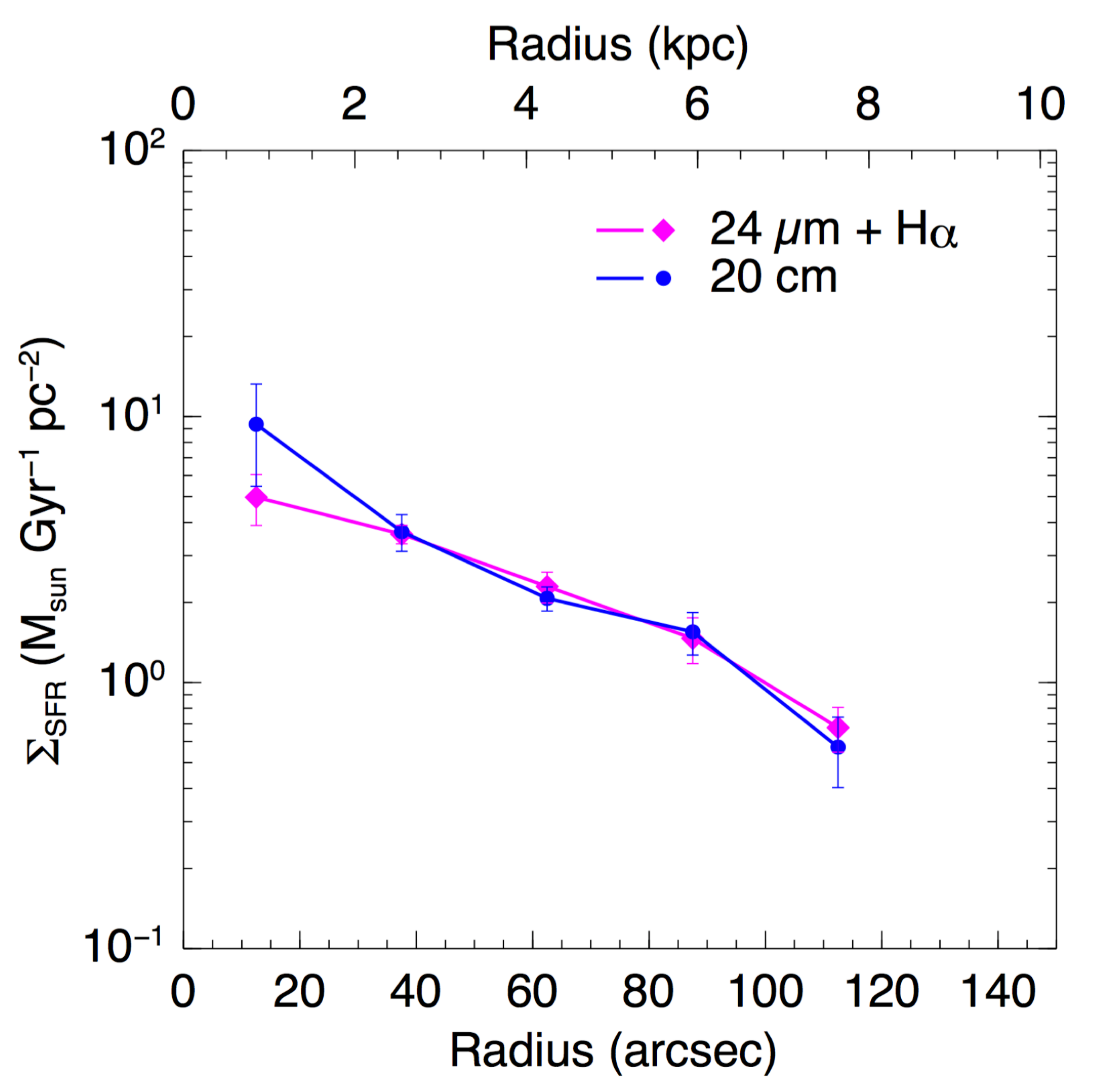}
\caption{
Top: radial surface density profiles of \Htwo\ (red squares), \HI\ (blue circles), SFR (magenta diamonds), and the total gas (black triangles). The vertical error bars represent the standard deviation of data in a radial bin. Bottom: SFR radial profiles from the linear combination of H$\alpha$ and 24 \um\ (magenta diamonds) and the radio continuum (blue circles). The vertical error bars are the standard deviation of the average in each bin of 25\ac. 
\label{fig:rprof}}
\end{center}
\end{figure}


\begin{figure*}
\begin{center}
\begin{tabular}{c@{\hspace{0.1in}}c@{\hspace{0.1in}}c@{\hspace{0.1in}}c}
\includegraphics[width=0.3\textwidth]{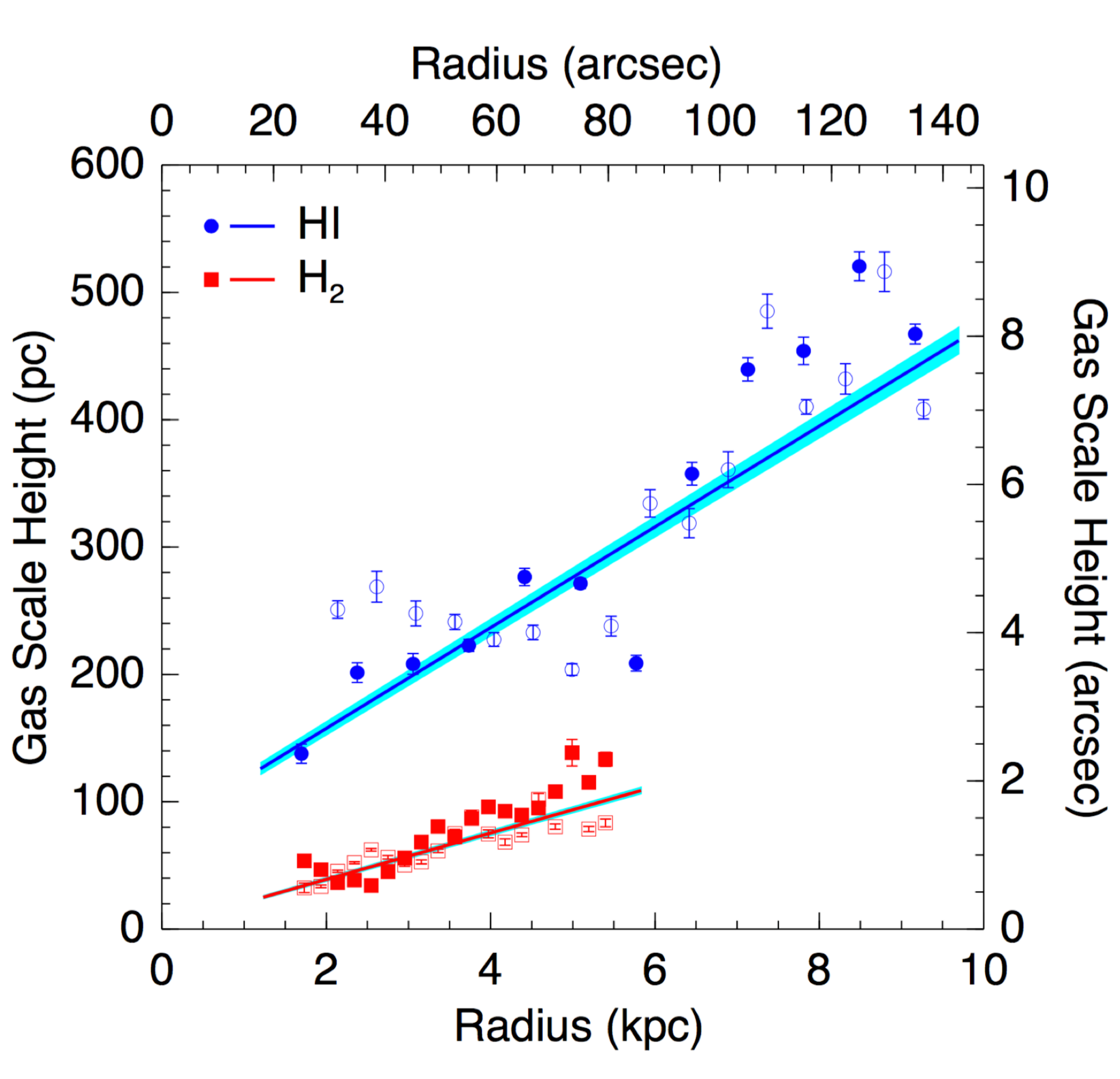}
\includegraphics[width=0.22\textwidth]{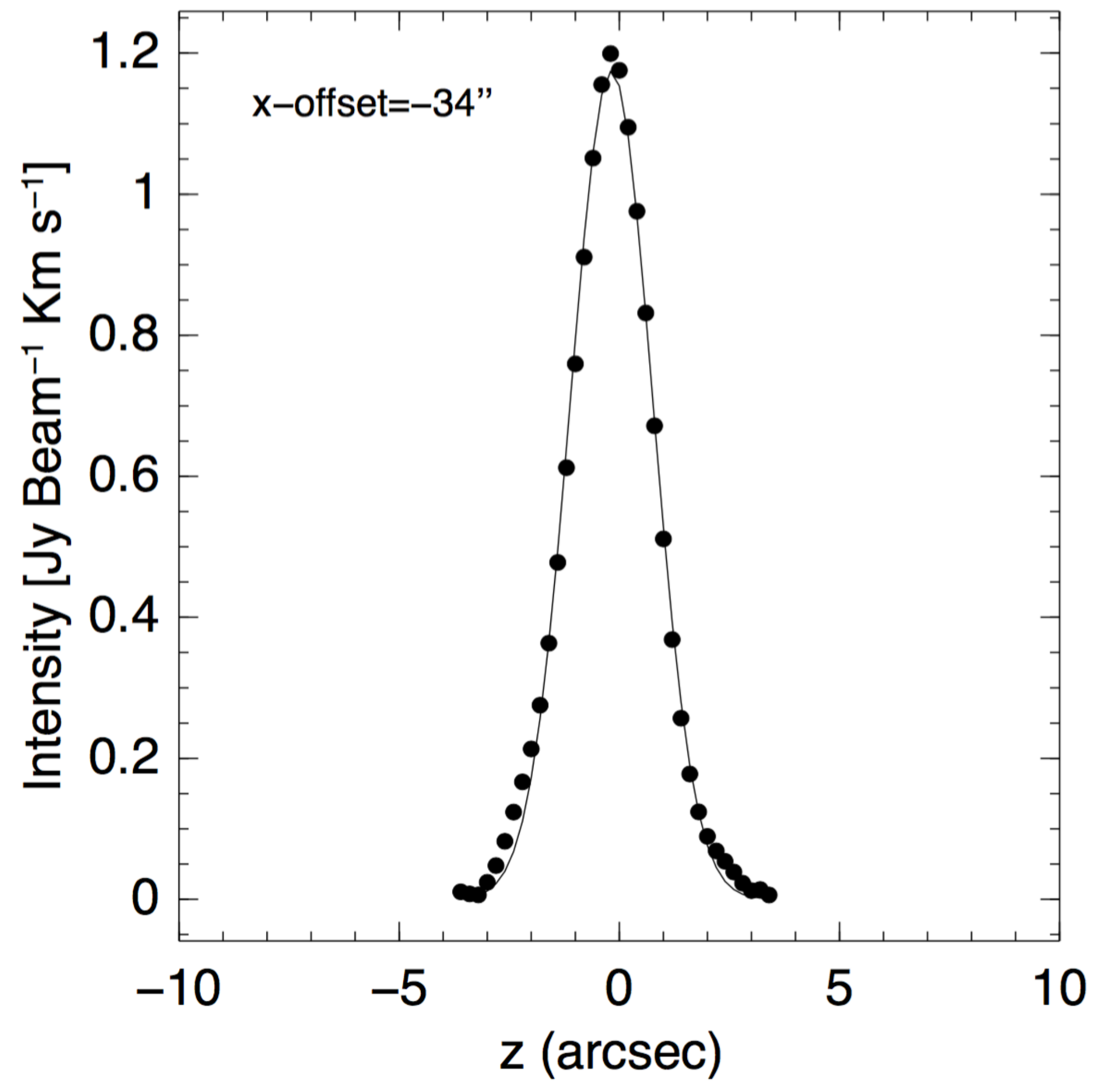}
\includegraphics[width=0.22\textwidth]{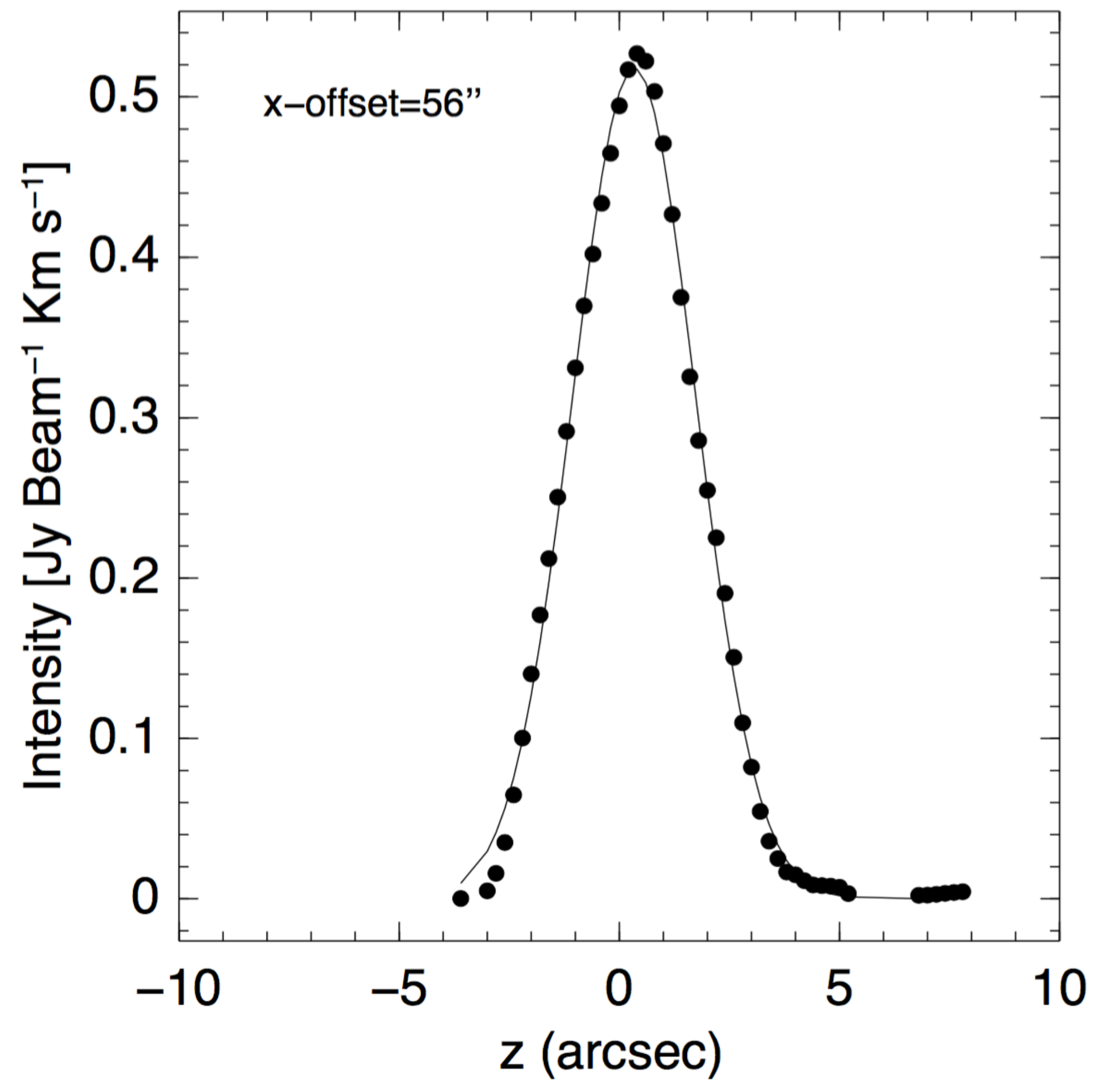}
\includegraphics[width=0.22\textwidth]{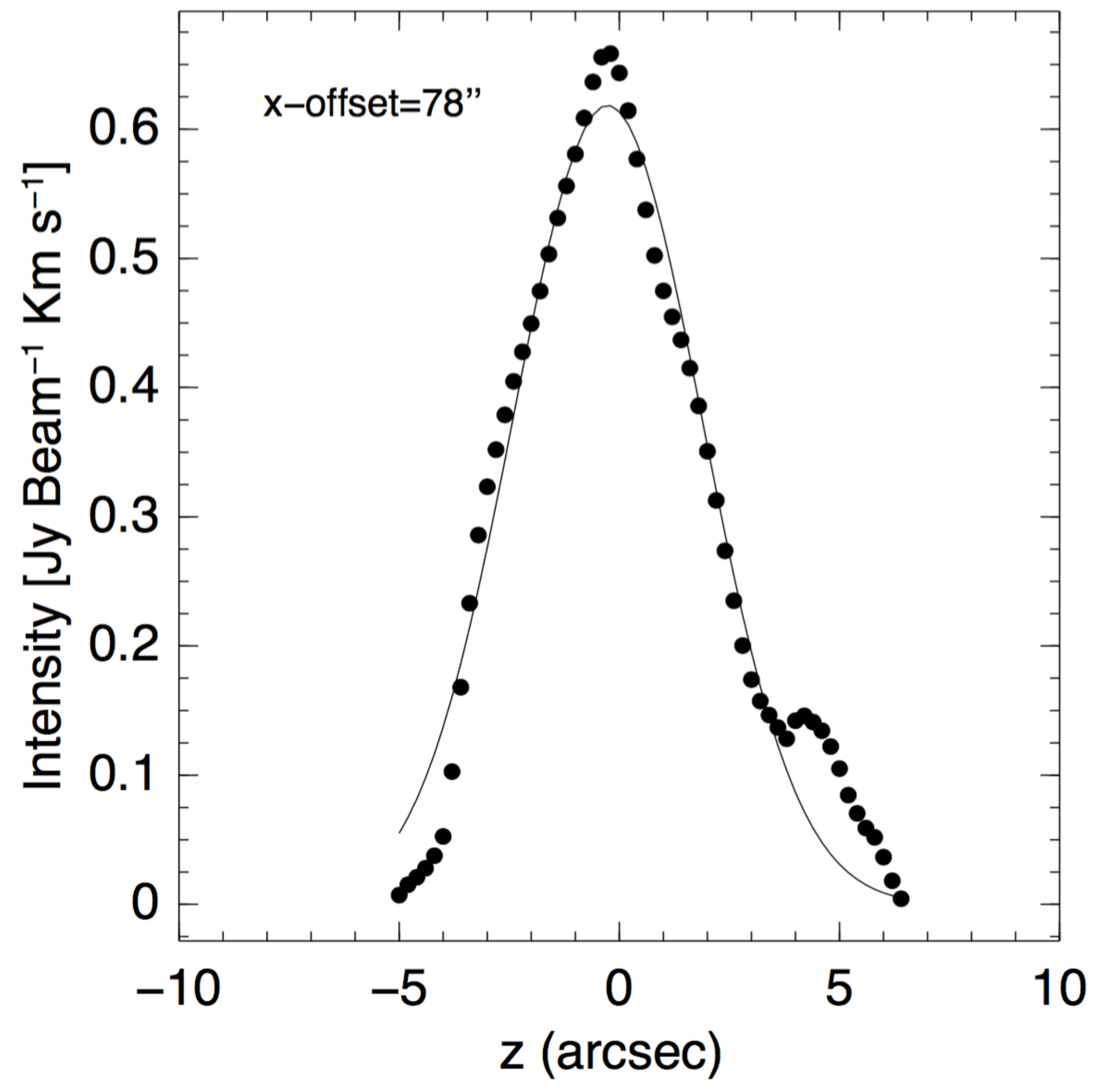}
\end{tabular}
\caption{Scale heights of \HI\ (blue circles) and CO (red squares) as functions of radius. The open (blueshifted side) and filled (redshifted side) circles are average values in bins of 3\ac\ (CO) and 7\ac\ (\HI). The lines are the linear  least-squares fit to the average data points. 
The shaded regions around the best fit lines show uncertainties of the best-fits and the vertical error bars show uncertainties of the Gaussian fitting for the scale height. The right three panels show Gaussian fits (line) to the vertical profiles (filled circles) of CO at $x$-offsets of $-$34\ac, 56\ac, and 78\ac. 
\label{fig:scaleh}}
\end{center}
\end{figure*}

We use the vertically integrated p-v diagrams in Figure \ref{fig:pvd} to obtain the radial profiles of \sighi\ and \sightwo\ using the PVD method, which is based on an assumption of circular rotation. Each pixel position in the p-v diagram allows us to derive a galactic radius and the face-on surface brightness at the radius:
\begin{equation}
R =  V_{\rm c} \left< \frac{x}{V_{\rm los}-V_{\rm sys}} \right> \quad\quad \rm{with}\ \it |V_{\rm los}-V_{\rm sys}| < V_{\rm c},
\label{eq_rad}
\end{equation}
where $V_{\rm c}$ is the assumed circular speed obtained from the rotation curve (bottom panel of Figure \ref{fig:pvd}) and $V_{\rm los}$ is the observed line-of-sight velocity at each position $x$. The angle brackets denote the mean value within a pixel. When we use the p-v diagram, we exclude the central regions ($|x| < 40$\ac\ and $|V_{\rm los}-V_{\rm sys}| < 40$ \kms) to avoid non-circular motions and confusion caused by the line of sight blending of emission from many different radii in the regions. Using the rotation curve, we find a radius corresponding to a position of $x$ and $V_{\rm los}$ in the p-v diagram. Finally, we convert the flux in the pixel into the face-on surface brightness and obtain the surface mass densities from the surface brightnesses using a standard Galactic value for the CO-to-\Htwo\ conversion factor (\citealt{1996A&A...308L..21S}; \citealt{2001ApJ...547..792D}) and assuming optically thin \HI\ emission:
\begin{equation}
\sightwo \, [\surm] = 3.2 I_{\rm CO} \,[\rm K \,\kms],
\label{xco}
\end{equation} 
\begin{equation}
\sighi \, [\surm] = 0.0146 I_{\rm HI} \,[\rm K \, \kms].
\label{xh1}
\end{equation}
We average the surface mass densities within each 10\ac\ bin and plot the average values of H$_2$ (red squares) and \HI\ (blue circles) as a function of radius in Figure \ref{fig:rprof} (top). The molecular and atomic gas densities include a factor of 1.36 for the helium correction. This figure shows that the molecular gas (\sightwo) is centrally concentrated while the atomic gas (\sighi) is almost uniformly distributed over the disk even beyond the CO disk. The surface density of the total gas (\siggas) is the combination of the molecular and atomic gas densities. The vertical error bars show the standard deviation of the average data within each bin. We also estimate the uncertainty from error propagation using the rms noise in the p-v diagram; this uncertainty is significantly smaller than the standard deviation. The uncertainties from the center outward are 0.14--0.01 \surm\ for \Htwo\ and 2.15--0.04 \surm\ for \HI. 
The horizontal error bars indicate the maximum and minimum radius obtained by using the angular and velocity resolutions in the calculation of radius (Equation \ref{eq_rad}). 

Since the circular speed could be affected by a bar, we do not exactly know the rotation curve near the center. Therefore, we also use a circular speed of 180 \kms\ for $V_{\rm c}$ instead of the rotation curve to examine how the radial profiles vary in the central regions within $\sim$40\ac, where the rotation curve does not match to the assumed flat velocity of 180 \kms. The maximum difference between the surface densities using the adopted rotation curve and the assumed flat rotation curve (180 \kms)  in the central regions is a factor of $\sim$2 for CO and $\sim$5 for \HI. However, the surface densities using the different $V_{\rm c}$ values are consistent with each other beyond $R\sim$40\ac. 
 
\subsection{Star Formation Rate}
We use a linear combination of H$\alpha$ and 24 \um\ (\citealt{2007ApJ...666..870C};  \citealt{2011AJ....142...37S}) to estimate the recent SFR density:
\begin{equation}
\sigsfr [\Msol \, {\rm kpc^{-2} \, yr^{-1}}] = 0.029\, I_{\rm H\alpha}
+ 0.0025\, I_{24\, \mu m},
\label{eq:sfrcombi}
\end{equation}
where the intensities of H$\alpha$ and 24 \um\ in units of MJy sr$^{-1}$ are obtained from RADPROF, which provides the radial profile in a face-on disk. The linear combination is meant to capture both obscured and unobscured star formation. \cite{2007ApJ...666..870C} derived the calibration from a sample of 33 nearby galaxies obtained by the SINGS ($Spitzer$ Infrared Nearby Galaxies Survey; \citealt{2003PASP..115..928K}) project.


Since the data do not include the velocity information, the PVD method is not usable to obtain the SFR  radial profile. Instead, we use the GIPSY task RADPROF, requiring integrated intensity strips and some input parameters (position angle, inclination, beam size, sigma, and initial estimate for the density distribution) and providing the radial surface density distribution based on the Lucy iterative scheme assuming axisymmetry \citep{1974AJ.....79..745L} developed by \citet{1988A&AS...72..427W}. 
The SFR surface density (\sigsfr) from RADPROF is shown as magenta diamonds in Figure \ref{fig:rprof} (top). The SFR density profile is dominated by the 24 \um\ term in Equation \ref{eq:sfrcombi}. The vertical error bars show the standard deviation of data in a bin of 10\ac. The SFR density is also centrally concentrated like the CO density, but the emission is extended beyond the CO disk, up to the \HI\ disk. We also derive another SFR estimate using the VLA 1.4 GHz radio continuum data \citep{2013A&A...553A.116V} based on the method given by \citet{2002A&A...385..412M}. We have compared the SFR radial profiles from the linear combination of H$\alpha$ and 24 \um\ and the radio continuum at the same resolution of 22\ac\ in Figure \ref{fig:rprof} (bottom). Overall, they are very consistent with each other except for the central regions within 40\ac, where the largest difference is a factor of 1.9. \citet{2018ApJ...853..128V} found that edge-on galaxies have a measurable optical depth at 24 \um\ and have lower flux ratios of 25/100 \um\ flux ratios compared to less inclined galaxies by a factor of 1.36.

\section{Gas Disk Thickness} \label{sec:disk}

NGC 4302 is the almost edge-on galaxy that allows us to measure the disk thickness  directly. First, we integrate the velocity channels over 950--980 \kms\ for the redshifted side and 1300--1330 \kms\ for the blueshifted side to obtain the terminal-velocity integrated intensity maps using the CO and \HI\ masked cubes. The radial ranges where the measurements are available are about 25\ac--80\ac\ for CO and about 25\ac--140\ac\ for \HI. Second, we fit a Gaussian function to vertical density profiles of the integrated intensity map at each $x$-offset (which approaches a galactic radius at the terminal velocity). Finally, we use the averaged Gaussian widths in radial bins of 3\ac\ for CO and 7\ac\ for \HI\ as the scale heights, where the beam size is deconvolved. Figure \ref{fig:scaleh} (left) shows the scale heights of \HI\ (blue circles) and CO  (red squares) as open (blueshifted side) and filled (redshifted side) circles with vertical error bars, representing the uncertainties of the Gaussian fits. The right three panels in the figure show the Gaussian fits to the vertical density profiles at different $x$-offsets.   
The uncertainty of the CO scale height is clearly smaller than that of the edge-on galaxy NGC 4013 ($i\sim90\degr$) using CARMA \citep{2020MNRAS.494.4558Y} thanks to the unprecedented sensitivity of ALMA. The blue and red lines are the linear least-squares fits to the average data points and the best fits for the scale heights ($h$),
\begin{equation}
    h_{\rm {HI}} =  (39.6 \pm 0.7) \times R~[\rm kpc] + (78.5 \pm 3.7) \quad [\rm{pc}],
\end{equation}

\begin{equation}
    h_{\rm H_2} = (18.2 \pm 0.3) \times R~[\rm kpc] + (2.6 \pm 0.9) \quad [\rm{pc}],
\end{equation}

will be used to derive the gas volume density.
The uncertainties of the best fits are shown as the shaded (cyan) region around the lines.  Both the CO and \HI\ scale heights increase as a function of radius, but the \HI\ gradient is much steeper than the CO gradient. NGC 4302 has a significantly steeper slope in the gas scale height compared to the other edge-on galaxies presented in the previous studies (\citealt{2014AJ....148..127Y}, \citeyear{2020MNRAS.494.4558Y}); steeper by a factor of $\sim$4 for CO and a factor of $\sim$2 for \HI\ compared to the average CO and \HI\ gradients of the other galaxies.  


\section{The Volumetric Star Formation Law}

We derive the mid-plane volume densities using the surface densities and the scale heights under assumptions of Gaussian (gas) and exponential (SFR) distributions along the vertical direction $z$. The surface densities are
\begin{equation}
\siggas = \int_{-\infty}^{+\infty}\rho_{\rm gas}e^{\frac{-z^2}{2h^2}} dz,
\sigsfr = \int_{-\infty}^{+\infty}\rho_{\rm SFR}e^{\frac{-|z|}{h}} dz.
\end{equation}
Therefore, the volume densities are
\begin{equation}
\rho_{\rm gas} = \frac{\siggas}{h_{\rm gas}\sqrt{2\pi}},  \qquad 
\rho_{\rm SFR} =  \frac{\sigsfr}{2 h_{\rm SFR}},
\label{eq_vol}
\end{equation}
where we use the CO scale height for $h_{\rm SFR}$ under the assumption that the molecular gas should have about the same scale height as the SF. The 24 \um\ scale height is likely not the SF scale height because of high-latitude dust being heated by a thinner SF layer.

\begin{figure*}
\begin{center}
\begin{tabular}{c@{\hspace{0.1in}}c}
\includegraphics[width=0.35\textwidth]{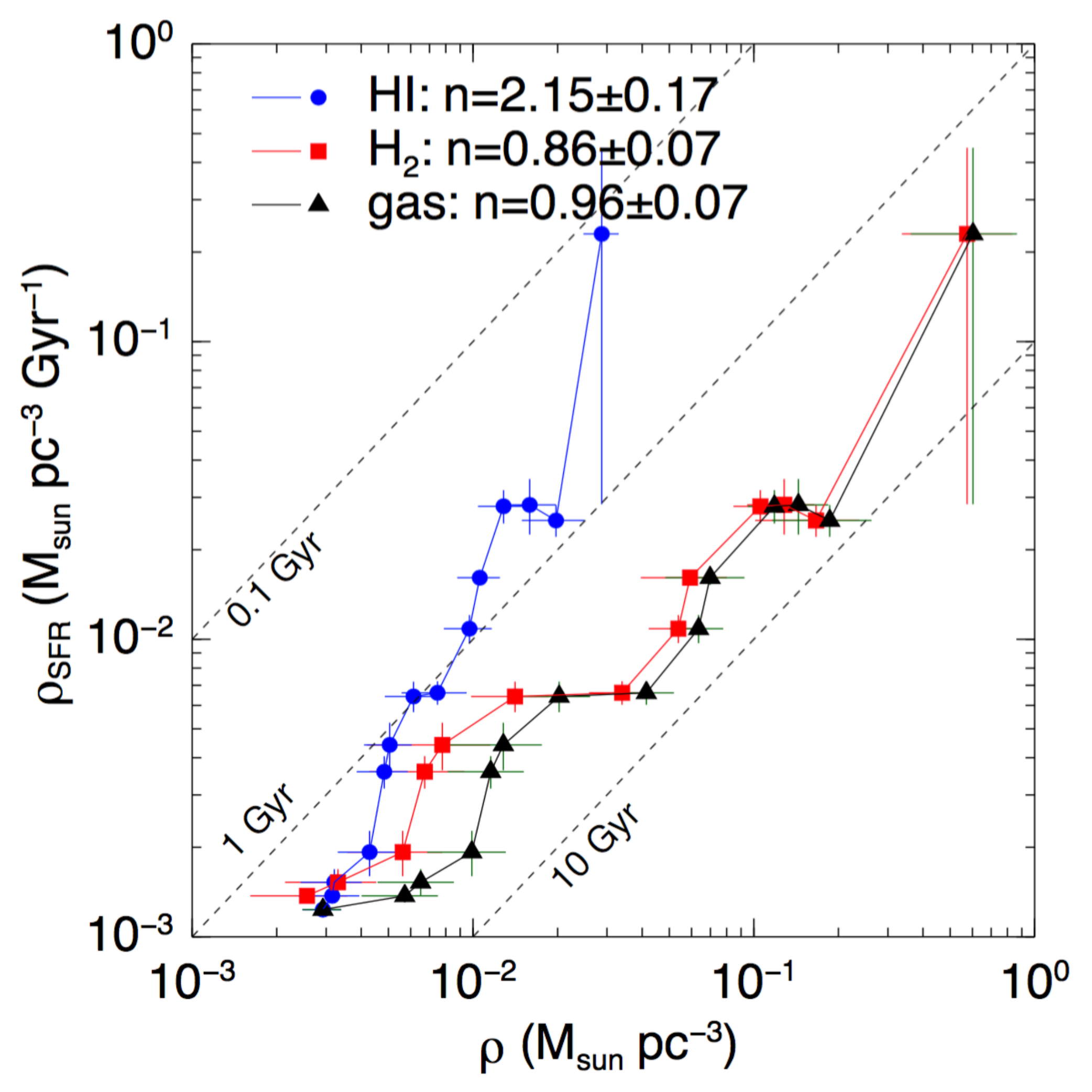}
\includegraphics[width=0.35\textwidth]{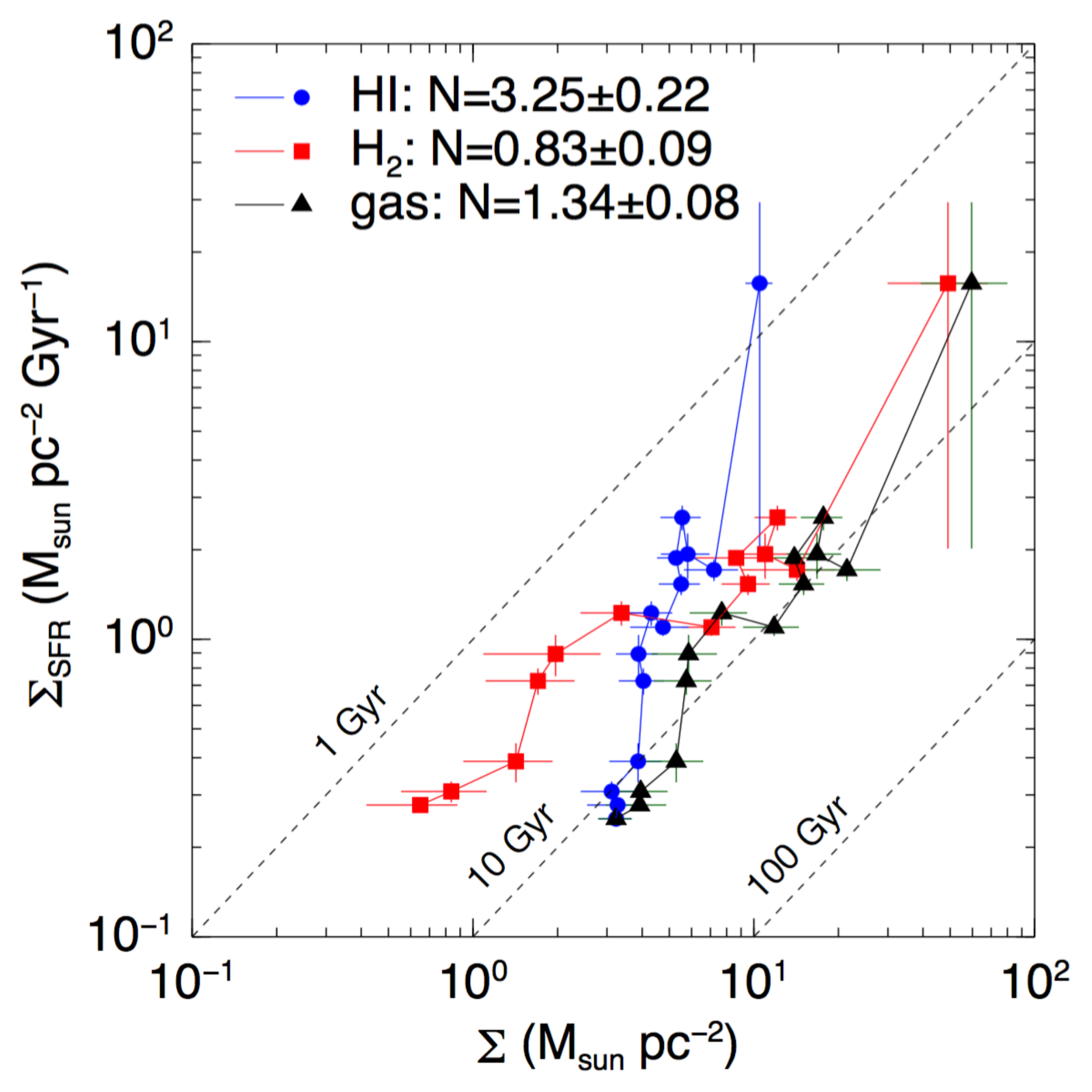}\\
\includegraphics[width=0.35\textwidth]{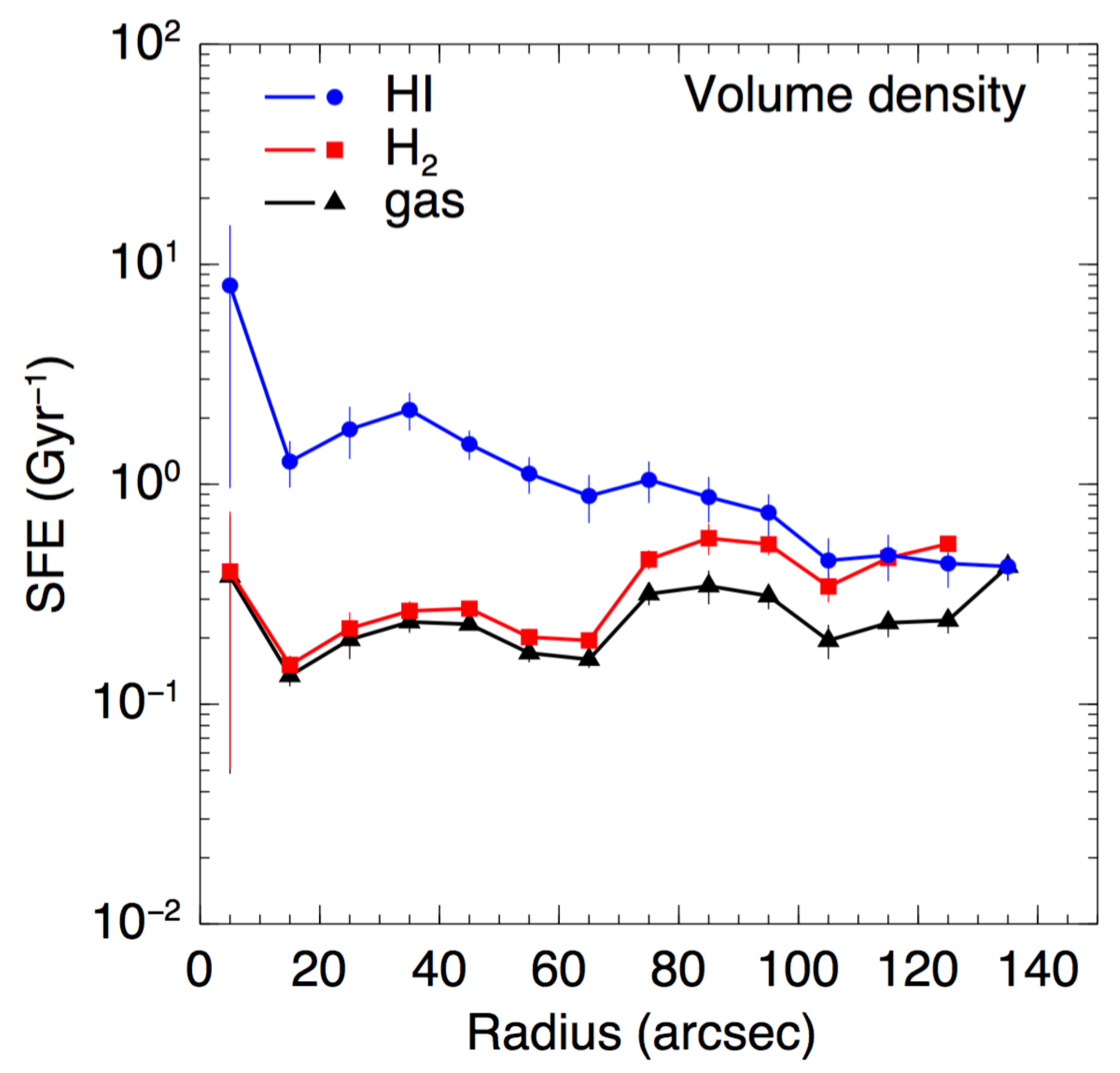}
\includegraphics[width=0.35\textwidth]{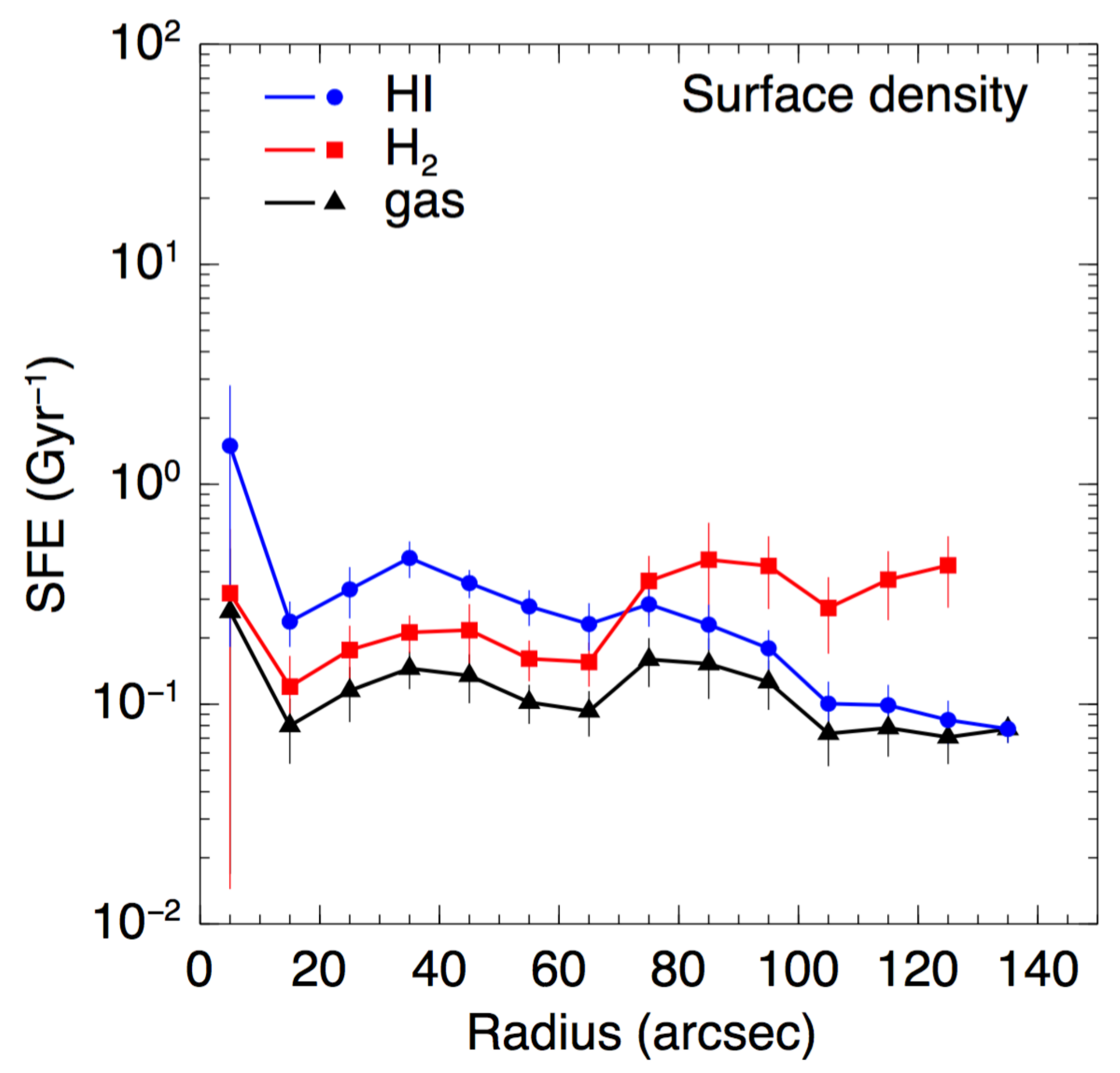}
\end{tabular}
\caption{Top panels: SFR volume density $\rho_{\rm SFR}$ as a function of  $\rho_{\rm HI}$ (blue circles), $\rho_{\rm H_2}$ (red squares), and $\rho_{\rm gas}$ (black triangles). SFR surface density \sigsfr\ as a function \sighi, \sightwo, and \siggas. The gas depletion time (SFE$^{-1}$) is labeled on the corresponding dashed lines. Bottom panels: \Htwo, \HI, and the total gas SFEs as a function of radius for volume (left) and surface (right) densities. The vertical error bars represent the standard deviation of data in each bin. 
\label{fig:sfl}}
\end{center}
\end{figure*}

We plot $\rho_{\rm SFR}$ against $\rho_{\rm HI}$, $\rho_{\rm H_2}$, and $\rho_{\rm gas}$ and \sigsfr\ against \sighi, \sightwo, and \siggas\ in the top panels of Figure \ref{fig:sfl} to investigate and compare the star formation laws based on volume and surface densities. We find tight power-law correlations between the SFR and \HI, \Htwo, and the total gas by fitting the ordinary least-squares bisector \citep{1990ApJ...364..104I} on a log-log scale:

\begin{equation}
\rhosfr \propto \rho_{\rm HI}^{2.15 \pm 0.17}, 
\rho_{\rm SFR} \propto \rho_{\rm H_2}^{0.86 \pm 0.07}, 
\rho_{\rm SFR} \propto \rho_{\rm gas}^{0.96 \pm 0.07},
\end{equation}
and 
\begin{equation}
\sigsfr \propto \Sigma_{\rm HI}^{3.25 \pm 0.22}, 
\sigsfr \propto \Sigma_{\rm H_2}^{0.83 \pm 0.09}, 
\sigsfr \propto \Sigma_{\rm gas}^{1.34 \pm 0.08}.
\end{equation}

The molecular power-law index of the volumetric SFL is comparable to that of the vertically integrated  SFL while the atomic and total gas indices of the volume density are noticeably flatter than the indices of the surface density. When we correct the SFR for the extinction in the central region ($R\leq$ 30\ac) by a factor of 1.9 that we obtained from the comparison between the radio continuum and the linear combination of H$\alpha$ and 24 \um, the both SFLs have a bit steeper slopes: 1.00 for the both molecular slopes, 1.09 (volume) and 1.57 (surface) for the total gas, and 2.44 (volume) and 3.82 (surface) for the atomic gas. 
The main reason for quite different indices between surface and volume densities for the atomic and total gas is that the \HI\ scale height increases with radius rapidly, causing a big difference in the gas surface and volume densities. 
In both cases, the \HI\ power-law index is obviously larger compared to the indices for the molecular and total gas since the \HI\ density is roughly constant until it decreases slowly in the outer regions, whereas the SFR density decreases rapidly with radius. On the other hand, a  tight correlation between SFR and \HI\ in volume and surface densities is not shown in many previous studies (e.g., \citealt{2008AJ....136.2846B}; \citealt{2011AJ....142...37S}; \citealt{2020MNRAS.494.4558Y}) though in a recent study, \citet{Bacchini_2019} present a strong correlation between $\rho_{\rm SFR}$ and $\rho_{\rm HI}$. 
The correlation shown in \HI, unlikely from the previous study of \citet{2020MNRAS.494.4558Y}, is caused by the different radial density distributions: this study using the derived rotation curve and the previous study assuming a flat rotation curve.


In terms of the star formation efficiency (SFE = $\rho_{\rm SFR}/\rho_{\rm gas}$ or \sigsfr/\siggas) in the bottom panels of Figure \ref{fig:sfl}, the SFE$_{\rm H_2}$ and SFE$_{\rm gas}$ appear to be roughly constant over the disk, while the atomic gas SFE$_{\rm HI}$ decreases as a function of radius in both volume and surface densities. The molecular gas depletion time (SFE$_{\rm H_2}^{-1}$) is in ranges of 1.8--6.6 Gyr (volume) and 2.2--8.3 Gyr (surface) and the total gas depletion time is in ranges of 2.4--7.4 Gyr (volume) and 3.8--14.1 Gyr (surface).



\section{Summary and Conclusion}
Using ALMA ($^{12}$CO $J=1\rightarrow0$) and VLA (\HI), we derive the molecular and atomic gas surface densities of the edge-on galaxy NGC 4302. We also estimate the SFR surface density from the linear combination of 24 \um\ and H$\alpha$ emissions. In order to obtain the mid-plane volume density, we measure the CO and \HI\ scale heights by fitting the Gaussian function to the vertical profiles and find that the scale heights increase significantly as a function of radius. Uncertainty of the CO scale height using ALMA is smaller than the uncertainty using CARMA given by \citet{2020MNRAS.494.4558Y}: the gradient uncertainty of the ALMA scale height is smaller by a factor of $\sim$7. 

Using the measured scale heights and surface densities, we infer the volume densities of CO, \HI, and SFR (24 \um\ $\pm$ H$\alpha$) to investigate the volumetric star formation law and compare the volumetric SFL with the vertically integrated SFL. we find strong power-law correlations between $\rho_{\rm HI}$, $\rho_{\rm H_2}$, $\rho_{\rm gas}$ and $\rho_{\rm SFR}$ with the indices of 2.15, 0.86, and 0.96, respectively. These volumetric power-law indices for \HI\ and the total gas (\HI+\Htwo) are quite smaller than the indices based on the surface densities while the \Htwo\ indices for both volume and surface densities are similar to each other. We also find that the SFE$_{\rm H_2}$ and SFE$_{\rm gas}$ are roughly constant, while the SFE$_{\rm HI}$ decreases as a function of radius in both volume and surface densities. 




\acknowledgments
We thank the anonymous referee for useful suggestions that improved this paper. 
K.Y. was supported by Basic Science Research Program through the National Research Foundation of Korea (NRF) funded by the Ministry of Education (grant number NRF-2021R1I1A1A01049891).
T.W. acknowledges financial support from the University of Illinois Vermilion River Fund for Astronomical Research.
This paper makes use of the following ALMA data: ADS/JAO.ALMA\#2019.1.00562.S. ALMA is a partnership of ESO (representing its member states), NSF (USA) and NINS (Japan), together with NRC (Canada), MOST and ASIAA (Taiwan), and KASI (Republic of Korea), in cooperation with the Republic of Chile. The Joint ALMA Observatory is operated by ESO, AUI/NRAO and NAOJ. The National Radio Astronomy Observatory is a facility of the National Science Foundation operated under cooperative agreement by Associated Universities, Inc.
 
%

\bibliography{refer}{}

\begin{thebibliography}{}
\expandafter\ifx\csname natexlab\endcsname\relax\def\natexlab#1{#1}\fi
\providecommand{\url}[1]{\href{#1}{#1}}
\providecommand{\dodoi}[1]{doi:~\href{http://doi.org/#1}{\nolinkurl{#1}}}
\providecommand{\doeprint}[1]{\href{http://ascl.net/#1}{\nolinkurl{http://ascl.net/#1}}}
\providecommand{\doarXiv}[1]{\href{https://arxiv.org/abs/#1}{\nolinkurl{https://arxiv.org/abs/#1}}}

\bibitem[{{Abramova} \& {Zasov}(2008)}]{2008ARep...52..257A}
{Abramova}, O.~V., \& {Zasov}, A.~V. 2008, Astronomy Reports, 52, 257,
  \dodoi{10.1134/S106377290804001X}

\bibitem[{{Aniano} {et~al.}(2011){Aniano}, {Draine}, {Gordon}, \&
  {Sandstrom}}]{2011PASP..123.1218A}
{Aniano}, G., {Draine}, B.~T., {Gordon}, K.~D., \& {Sandstrom}, K. 2011, \pasp,
  123, 1218, \dodoi{10.1086/662219}

\bibitem[{{Athanassoula} \& {Bureau}(1999)}]{1999ApJ...522..699A}
{Athanassoula}, E., \& {Bureau}, M. 1999, \apj, 522, 699,
  \dodoi{10.1086/307677}

\bibitem[{{Bacchini} {et~al.}(2020){Bacchini}, {Fraternali}, {Pezzulli}, \&
  {Marasco}}]{2020A&A...644A.125B}
{Bacchini}, C., {Fraternali}, F., {Pezzulli}, G., \& {Marasco}, A. 2020, \aap,
  644, A125, \dodoi{10.1051/0004-6361/202038962}

\bibitem[{{Bacchini} {et~al.}(2019){Bacchini}, {Fraternali, Filippo}, {Iorio,
  Giuliano}, \& {Pezzulli, Gabriele}}]{Bacchini_2019}
{Bacchini}, C., C., {Fraternali, Filippo}, {Iorio, Giuliano}, \& {Pezzulli,
  Gabriele}. 2019, A\&A, 622, A64, \dodoi{10.1051/0004-6361/201834382}

\bibitem[{{Bigiel} {et~al.}(2008){Bigiel}, {Leroy}, {Walter}, {Brinks}, {de
  Blok}, {Madore}, \& {Thornley}}]{2008AJ....136.2846B}
{Bigiel}, F., {Leroy}, A., {Walter}, F., {et~al.} 2008, \aj, 136, 2846,
  \dodoi{10.1088/0004-6256/136/6/2846}

\bibitem[{{Blitz} \& {Rosolowsky}(2006)}]{2006ApJ...650..933B}
{Blitz}, L., \& {Rosolowsky}, E. 2006, \apj, 650, 933, \dodoi{10.1086/505417}

\bibitem[{{Calzetti} {et~al.}(2007){Calzetti}, {Kennicutt}, {Engelbracht},
  {Leitherer}, {Draine}, {Kewley}, {Moustakas}, {Sosey}, {Dale}, {Gordon},
  {Helou}, {Hollenbach}, {Armus}, {Bendo}, {Bot}, {Buckalew}, {Jarrett}, {Li},
  {Meyer}, {Murphy}, {Prescott}, {Regan}, {Rieke}, {Roussel}, {Sheth}, {Smith},
  {Thornley}, \& {Walter}}]{2007ApJ...666..870C}
{Calzetti}, D., {Kennicutt}, R.~C., {Engelbracht}, C.~W., {et~al.} 2007, \apj,
  666, 870, \dodoi{10.1086/520082}

\bibitem[{Chung {et~al.}(2009)Chung, van Gorkom, Kenney, Crowl, \&
  Vollmer}]{Chung_2009}
Chung, A., van Gorkom, J.~H., Kenney, J. D.~P., Crowl, H., \& Vollmer, B. 2009,
  The Astronomical Journal, 138, 1741, \dodoi{10.1088/0004-6256/138/6/1741}

\bibitem[{Chung {et~al.}(2007)Chung, van Gorkom, Kenney, \&
  Vollmer}]{Chung_2007}
Chung, A., van Gorkom, J.~H., Kenney, J. D.~P., \& Vollmer, B. 2007, The
  Astrophysical Journal, 659, L115, \dodoi{10.1086/518034}

\bibitem[{{Dame} {et~al.}(2001){Dame}, {Hartmann}, \&
  {Thaddeus}}]{2001ApJ...547..792D}
{Dame}, T.~M., {Hartmann}, D., \& {Thaddeus}, P. 2001, \apj, 547, 792,
  \dodoi{10.1086/318388}

\bibitem[{{Giovanelli} {et~al.}(2007){Giovanelli}, {Haynes}, {Kent},
  {Saintonge}, {Stierwalt}, {Altaf}, {Balonek}, {Brosch}, {Brown}, {Catinella},
  {Furniss}, {Goldstein}, {Hoffman}, {Koopmann}, {Kornreich}, {Mahmood},
  {Martin}, {Masters}, {Mitschang}, {Momjian}, {Nair}, {Rosenberg}, \&
  {Walsh}}]{2007AJ....133.2569G}
{Giovanelli}, R., {Haynes}, M.~P., {Kent}, B.~R., {et~al.} 2007, \aj, 133,
  2569, \dodoi{10.1086/516635}

\bibitem[{Heald {et~al.}(2007)Heald, Rand, Benjamin, \& Bershady}]{Heald_2007}
Heald, G.~H., Rand, R.~J., Benjamin, R.~A., \& Bershady, M.~A. 2007, The
  Astrophysical Journal, 663, 933, \dodoi{10.1086/518087}

\bibitem[{Howk \& Savage(1999)}]{Howk_1999}
Howk, J.~C., \& Savage, B.~D. 1999, \aj, 117, 2077, \dodoi{10.1086/300857}

\bibitem[{{Isobe} {et~al.}(1990){Isobe}, {Feigelson}, {Akritas}, \&
  {Babu}}]{1990ApJ...364..104I}
{Isobe}, T., {Feigelson}, E.~D., {Akritas}, M.~G., \& {Babu}, G.~J. 1990, \apj,
  364, 104, \dodoi{10.1086/169390}

\bibitem[{{Kennicutt} {et~al.}(2003){Kennicutt}, {Armus}, {Bendo}, {Calzetti},
  {Dale}, {Draine}, {Engelbracht}, {Gordon}, {Grauer}, {Helou}, {Hollenbach},
  {Jarrett}, {Kewley}, {Leitherer}, {Li}, {Malhotra}, {Regan}, {Rieke},
  {Rieke}, {Roussel}, {Smith}, {Thornley}, \& {Walter}}]{2003PASP..115..928K}
{Kennicutt}, R.~C., {Armus}, L., {Bendo}, G., {et~al.} 2003, \pasp, 115, 928,
  \dodoi{10.1086/376941}

\bibitem[{{Kennicutt}(1998)}]{1998ApJ...498..541K}
{Kennicutt}, Jr., R.~C. 1998, \apj, 498, 541, \dodoi{10.1086/305588}

\bibitem[{Komugi {et~al.}(2008)Komugi, Sofue, Kohno, Nakanishi, Onodera, Egusa,
  \& Muraoka}]{Komugi_2008}
Komugi, S., Sofue, Y., Kohno, K., {et~al.} 2008, The Astrophysical Journal
  Supplement Series, 178, 225, \dodoi{10.1086/590469}

\bibitem[{{Krumholz} {et~al.}(2012){Krumholz}, {Dekel}, \&
  {McKee}}]{2012ApJ...745...69K}
{Krumholz}, M.~R., {Dekel}, A., \& {McKee}, C.~F. 2012, \apj, 745, 69,
  \dodoi{10.1088/0004-637X/745/1/69}

\bibitem[{{Lucy}(1974)}]{1974AJ.....79..745L}
{Lucy}, L.~B. 1974, \aj, 79, 745, \dodoi{10.1086/111605}

\bibitem[{{Makarov} {et~al.}(2014){Makarov}, {Prugniel}, {Terekhova},
  {Courtois}, \& {Vauglin}}]{2014A&A...570A..13M}
{Makarov}, D., {Prugniel}, P., {Terekhova}, N., {Courtois}, H., \& {Vauglin},
  I. 2014, \aap, 570, A13, \dodoi{10.1051/0004-6361/201423496}

\bibitem[{{Marasco} {et~al.}(2017){Marasco}, {Fraternali}, {van der Hulst}, \&
  {Oosterloo}}]{2017A&A...607A.106M}
{Marasco}, A., {Fraternali}, F., {van der Hulst}, J.~M., \& {Oosterloo}, T.
  2017, \aap, 607, A106, \dodoi{10.1051/0004-6361/201731054}

\bibitem[{{McMullin} {et~al.}(2007){McMullin}, {Waters}, {Schiebel}, {Young},
  \& {Golap}}]{2007ASPC..376..127M}
{McMullin}, J.~P., {Waters}, B., {Schiebel}, D., {Young}, W., \& {Golap}, K.
  2007, in Astronomical Society of the Pacific Conference Series, Vol. 376,
  Astronomical Data Analysis Software and Systems XVI, ed. {R.~A.~Shaw,
  F.~Hill, \& D.~J.~Bell}, 127

\bibitem[{{Mogotsi} {et~al.}(2016){Mogotsi}, {de Blok}, {Cald{\'u}-Primo},
  {Walter}, {Ianjamasimanana}, \& {Leroy}}]{2016AJ....151...15M}
{Mogotsi}, K.~M., {de Blok}, W.~J.~G., {Cald{\'u}-Primo}, A., {et~al.} 2016,
  \aj, 151, 15, \dodoi{10.3847/0004-6256/151/1/15}

\bibitem[{{Mould} {et~al.}(2000){Mould}, {Huchra}, {Freedman}, {Kennicutt},
  {Ferrarese}, {Ford}, {Gibson}, {Graham}, {Hughes}, {Illingworth}, {Kelson},
  {Macri}, {Madore}, {Sakai}, {Sebo}, {Silbermann}, \&
  {Stetson}}]{2000ApJ...529..786M}
{Mould}, J.~R., {Huchra}, J.~P., {Freedman}, W.~L., {et~al.} 2000, \apj, 529,
  786, \dodoi{10.1086/308304}

\bibitem[{{Murgia} {et~al.}(2002){Murgia}, {Crapsi}, {Moscadelli}, \&
  {Gregorini}}]{2002A&A...385..412M}
{Murgia}, M., {Crapsi}, A., {Moscadelli}, L., \& {Gregorini}, L. 2002, \aap,
  385, 412, \dodoi{10.1051/0004-6361:20020140}

\bibitem[{{Rand}(1996)}]{1996ApJ...462..712R}
{Rand}, R.~J. 1996, \apj, 462, 712, \dodoi{10.1086/177184}

\bibitem[{{Schmidt}(1959)}]{1959ApJ...129..243S}
{Schmidt}, M. 1959, \apj, 129, 243, \dodoi{10.1086/146614}

\bibitem[{{Schruba} {et~al.}(2011){Schruba}, {Leroy}, {Walter}, {Bigiel},
  {Brinks}, {de Blok}, {Dumas}, {Kramer}, {Rosolowsky}, {Sandstrom},
  {Schuster}, {Usero}, {Weiss}, \& {Wiesemeyer}}]{2011AJ....142...37S}
{Schruba}, A., {Leroy}, A.~K., {Walter}, F., {et~al.} 2011, \aj, 142, 37,
  \dodoi{10.1088/0004-6256/142/2/37}

\bibitem[{{Shostak} \& {van der Kruit}(1984)}]{1984A&A...132...20S}
{Shostak}, G.~S., \& {van der Kruit}, P.~C. 1984, \aap, 132, 20

\bibitem[{{Sofue} \& {Rubin}(2001)}]{2001ARA&A..39..137S}
{Sofue}, Y., \& {Rubin}, V. 2001, \araa, 39, 137,
  \dodoi{10.1146/annurev.astro.39.1.137}

\bibitem[{{Strong} \& {Mattox}(1996)}]{1996A&A...308L..21S}
{Strong}, A.~W., \& {Mattox}, J.~R. 1996, \aap, 308, L21

\bibitem[{{van der Hulst} {et~al.}(1992){van der Hulst}, {Terlouw}, {Begeman},
  {Zwitser}, \& {Roelfsema}}]{1992ASPC...25..131V}
{van der Hulst}, J.~M., {Terlouw}, J.~P., {Begeman}, K.~G., {Zwitser}, W., \&
  {Roelfsema}, P.~R. 1992, in Astronomical Society of the Pacific Conference
  Series, Vol.~25, Astronomical Data Analysis Software and Systems I, ed. D.~M.
  {Worrall}, C.~{Biemesderfer}, \& J.~{Barnes}, 131

\bibitem[{{Vargas} {et~al.}(2018){Vargas}, {Mora-Partiarroyo}, {Schmidt},
  {Rand}, {Stein}, {Walterbos}, {Wang}, {Basu}, {Patterson}, {Kepley}, {Beck},
  {Irwin}, {Heald}, {Li}, \& {Wiegert}}]{2018ApJ...853..128V}
{Vargas}, C.~J., {Mora-Partiarroyo}, S.~C., {Schmidt}, P., {et~al.} 2018, \apj,
  853, 128, \dodoi{10.3847/1538-4357/aaa47f}

\bibitem[{{Vollmer} {et~al.}(2013){Vollmer}, {Soida}, {Beck}, {Chung},
  {Urbanik}, {Chy{\.z}y}, {Otmianowska-Mazur}, \&
  {Kenney}}]{2013A&A...553A.116V}
{Vollmer}, B., {Soida}, M., {Beck}, R., {et~al.} 2013, \aap, 553, A116,
  \dodoi{10.1051/0004-6361/201321163}

\bibitem[{{Warmels}(1988)}]{1988A&AS...72..427W}
{Warmels}, R.~H. 1988, \aaps, 72, 427

\bibitem[{{Wilson} {et~al.}(2011){Wilson}, {Warren}, {Irwin}, {Knapen},
  {Israel}, {Serjeant}, {Attewell}, {Bendo}, {Brinks}, \&
  {Butner}}]{2011MNRAS.410.1409W}
{Wilson}, C.~D., {Warren}, B.~E., {Irwin}, J., {et~al.} 2011, \mnras, 410,
  1409, \dodoi{10.1111/j.1365-2966.2010.17646.x}

\bibitem[{{Wong} \& {Blitz}(2002)}]{2002ApJ...569..157W}
{Wong}, T., \& {Blitz}, L. 2002, \apj, 569, 157, \dodoi{10.1086/339287}

\bibitem[{{Yim} \& {van der Hulst}(2016)}]{2016MNRAS.463.2092Y}
{Yim}, K., \& {van der Hulst}, J.~M. 2016, \mnras, 463, 2092,
  \dodoi{10.1093/mnras/stw2118}

\bibitem[{{Yim} {et~al.}(2011){Yim}, {Wong}, {Howk}, \& {van der
  Hulst}}]{2011AJ....141...48Y}
{Yim}, K., {Wong}, T., {Howk}, J.~C., \& {van der Hulst}, J.~M. 2011, \aj, 141,
  48, \dodoi{10.1088/0004-6256/141/2/48}

\bibitem[{{Yim} {et~al.}(2020){Yim}, {Wong}, {Rand}, \&
  {Schinnerer}}]{2020MNRAS.494.4558Y}
{Yim}, K., {Wong}, T., {Rand}, R.~J., \& {Schinnerer}, E. 2020, \mnras, 494,
  4558, \dodoi{10.1093/mnras/staa1020}

\bibitem[{{Yim} {et~al.}(2014){Yim}, {Wong}, {Xue}, {Rand}, {Rosolowsky}, {van
  der Hulst}, {Benjamin}, \& {Murphy}}]{2014AJ....148..127Y}
{Yim}, K., {Wong}, T., {Xue}, R., {et~al.} 2014, \aj, 148, 127,
  \dodoi{10.1088/0004-6256/148/6/127}

\bibitem[{Zschaechner {et~al.}(2015)Zschaechner, Rand, \&
  Walterbos}]{Zschaechner_2015}
Zschaechner, L.~K., Rand, R.~J., \& Walterbos, R. 2015, The Astrophysical
  Journal, 799, 61, \dodoi{10.1088/0004-637x/799/1/61}

\end{thebibliography}
\bibliographystyle{aasjournal}

\end{document}